\documentclass[prb, twocolumn, showpacs, english, preprintnumbers, amsmath, amssymb, superscriptaddress, aps]{revtex4-1}
\usepackage{latexsym}
\usepackage{epsfig,psfrag}
\usepackage{dcolumn}
\usepackage{bm}
\usepackage{graphicx}
\usepackage{color}
\usepackage{esint}
\usepackage{babel}
\usepackage{amsfonts}
\usepackage{enumerate}
\begin{document}
\title{Tunneling spectroscopy of Majorana-Kondo devices} 

\author{Erik Eriksson}
\affiliation{Institut f\"ur Theoretische Physik,
 Heinrich-Heine-Universit\"at, D-40225 D\"usseldorf, Germany}
\author{Andrea Nava}
\affiliation{Dipartimento di Fisica, Universit\`a della Calabria and INFN - Gruppo collegato di Cosenza, I-87036 Arcavacata di Rende, Cosenza, Italy}
\author{Christophe Mora}
\affiliation{Laboratoire Pierre Aigrain, {\'E}cole Normale Sup{\'e}rieure,
Universit{\'e} Paris 7 Diderot, CNRS;\\ 24 rue Lhomond, F-75005 Paris, France}
\author{Reinhold Egger$^1$}

\begin{abstract}
 
We study the local density of states (LDOS) in systems of Luttinger-liquid nanowires connected to a common mesoscopic superconducting island, in which Majorana bound states give rise to different types of topological Kondo effects. We show that electron interactions enhance the low-energy LDOS in the leads close to the island, with unusual exponents due to Kondo physics that can be probed in tunneling experiments.   
 
\end{abstract}

\maketitle

\section{Introduction}

Majorana bound states have become of major interest in condensed matter physics, \cite{mbsrev1,mbsrev2,mbsrev3,exp1,exp2,exp3,exp4,exp5,exp6} due to potential applications as building blocks in fault-tolerant quantum computing \cite{kitaev2003} and the possibility to engineer such topological states using conventional $s$-wave superconductors and spin-orbit coupling. \cite{FuKane,sao,ORO} Information in these states is encoded non-locally, with the long-range entanglement providing a mechanism for electron teleportation. \cite{fu2010}

Recently, it has been realized that the topologically protected ground-state subspace formed by several Majorana bound states can act as a non-local quantum impurity, which when subjected to strong charging effects and coupled to conduction electrons can give rise to a {\em topological Kondo effect}. \cite{beri1} Here a stable non-Fermi liquid behavior is obtained, reminiscent of the multichannel Kondo effect but robust against perturbations. In Ref.~\onlinecite{nrg} the full crossover was studied using numerical renormalization group. The situation with an arbitrary number of leads of interacting electrons was studied in Refs.~\onlinecite{beri2,altland1}, where in addition an interaction-induced intermediate-coupling unstable fixed point was discovered. The topological protection of this novel Kondo effect opens new possibilities for the experimental observation of multi-channel Kondo impurity dynamics. \cite{abet1,abet2} Additional physical effects can be observed when including a Josephson coupling to the mesoscopic island hosting the Majorana bound states; phase fluctuations then cause a non-trivial interplay between topological Kondo and resonant Andreev reflection processes, giving a continuous manifold of stable non-Fermi liquid states. \cite{emze} With $N$ wires each connected to one Majorana on the island, the symmetry group of this topological Kondo effect is SO$_1$($N$), previously encountered also for a junction of Ising chains, \cite{tsvelik} unlike that of Ref.~\onlinecite{beri1} which is SO$_2$($N$). 

The search for observable predictions regarding the topological Kondo effect has so far been focused on charge transport through the system \cite{beri1,beri2,altland1,zae,abet1,emze} or measurements of the occupation of pairs of Majorana zero modes, analogous to magnetization. \cite{abet1} In this paper, we show that the local density of states (LDOS) of the lead electrons close to the island provide a clear signature of the topological Kondo effect of B\'eri and Cooper, \cite{beri1} directly measurable with a scanning tunneling microscope (STM). In particular, we show that the LDOS close to the island follows the power law $ \rho(\omega) \sim \omega^{\frac{1}{NK} + \frac{N-1}{N} K-1}$ as a function of energy $\omega \to 0$, where $K$ is the Luttinger liquid parameter for the electron-electron interaction strength with $K=1$ for non-interacting leads and $K<1$ for repulsive interactions. Hence for realistic values $1/(N-1) < K <1$, we have a diverging LDOS in the zero-bias limit close to the junction. 

In contrast to the usual picture of a power-law vanishing of the low-energy LDOS in a Luttinger liquid with or without boundary/impurity, \cite{gogolinbook,KF,delft,eggert,kje} an interaction-induced divergence is in fact a rather generic feature of Luttinger-liquid wire junctions, \cite{adrs} and Luttinger-liquid  junctions with a superconductor, with \cite{Fidkowski} or without \cite{winkelholz,LL} Majorana bound states. The key feature of the SO$_2$($N$) topological Kondo effect of Ref.~\onlinecite{beri1} is that the power law governing the divergence depends on the number $N$ of leads participating in the effect, making adjustable gate voltages a route to observe this signature. This $N$ dependence of the LDOS is however absent in the SO$_1$($N$) topological Kondo effect of Ref.~\onlinecite{emze}, where we find the zero-energy divergence  $ \rho(\omega) \sim \omega^{ K-1}$ for all fixed points within the non-Fermi liquid manifold, which is the same power law as that encountered for perfect Andreev reflection at a single Luttinger-liquid junction with a Majorana fermion. \cite{Fidkowski}

The paper is organized as follows. In Sec.~\ref{model}, we review the device under study and the emerging low-energy theories, found in Refs.~\onlinecite{beri1,beri2,altland1,emze}. In Secs.~\ref{Green} and \ref{sec:ldos} we show how methods \cite{adrs} for calculating the LDOS in Luttinger-liquid wire junctions can be applied to our models, and in Sec.~\ref{sec:ldos2} we derive the results of this paper, computing the LDOS in the topological Kondo model. Unless stated otherwise, we use units such that $\hbar =1$.

\section{Model} \label{model}

\subsection{Device setup}

\begin{figure}[t!]
\centering
\includegraphics[width=8cm]{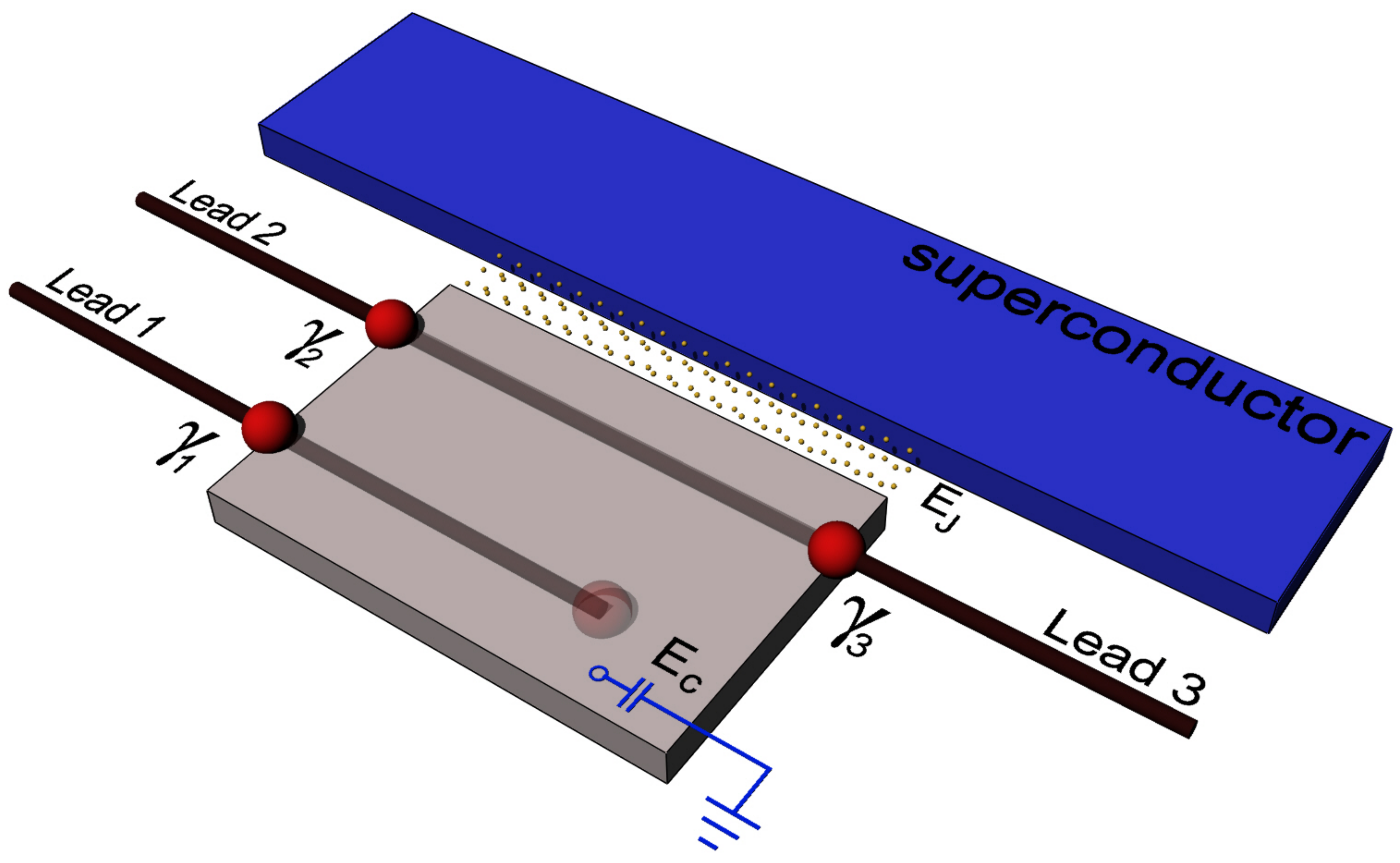}
\caption{\label{fig:1} 
(Color online) Schematic setup for a Majorana device hosting the topological Kondo effect. Spin-orbit coupled semiconductor nanowires (two in the figure) are deposited on top of an ordinary superconducing island (grey box) with charging energy $E_C$. In a magnetic field, Majorana bound states $\gamma_i$ (red dots) are formed at the ends of the wire parts coupled to the superconductor (dark grey). Gate voltages create tunnel barriers between $N$ Majorana fermions and the $N$ normal leads (in the figure, $N=3$). This leads to an SO$_2$($N$) topological Kondo effect at low temperature.\cite{beri1} When Josephson coupling the superconducting island to an additional bulk superconductor (blue), the system will, in the limit of large Josephson energy $E_J$, give an SO$_1$($N$) topological Kondo effect which becomes tunable by the lead-Majorana couplings.\cite{emze} }
\end{figure}

We consider the setup where the topological Kondo effect can take place,\cite{beri1} namely a mesoscopic s-wave superconducting island hosting a set of $N_{\textrm{tot}}$ localized Majorana bound states, of which $N \geq 3$ are tunnel-coupled to normal leads of conduction electrons. This setup is sketched in Fig.~\ref{fig:1}. Experimentally, this can be achieved by depositing $N_{\textrm{tot}}/2$ nanowires with strong spin-orbit coupling, e.g. InSb or InAs, subjected to a magnetic field, on top of a floating mesoscopic superconducting island; this creates $N_{\textrm{tot}}$ Majorana bound states, one at each end of the wire parts that are on top of the superconductor. \cite{exp1,exp2,exp3,exp4,exp5,exp6,mbsrev1,mbsrev2,mbsrev3} With proper gating, $N$ of these $N_{\textrm{tot}}$ Majoranas are tunnel coupled to the $N$ normal parts of the nanowires, which then act as leads. We will also consider a generalized setup, where the island is Josephson coupled to a bulk s-wave superconductor. \cite{emze}

The full Hamiltonian of the system under consideration is hence given by $H = H_{\textrm{leads}} + H_{\textrm{island}} + H_t$.

The normal leads of effectively spinless electrons are described by the Hamiltonian 
\begin{equation}
H_{\textrm{leads}} = - \mathrm{i} v_F \sum_{j=1}^N  \int_0^{\infty} \mathrm{d}x \ \left[ \Psi^{\dagger}_{j,O} \partial_x \Psi_{j,O} - \Psi^{\dagger}_{j,I} \partial_x \Psi_{j,I} \right] ,
\end{equation}
with fermionic fields $\Psi_j(x)$ for each lead $j$, consisting of outgoing ($O$) and incoming ($I$) components (i.e.~right and left movers). We assume all leads are identical. At $x=0$, we have the boundary condition $\Psi_{j,O}(0) = \Psi_{j,I}(0) \equiv \Psi_{j}(0)$ for disconnected leads. However, here the lead electrons are coupled to the localized Majorana modes on the island. These are described by operators $\gamma_j$ obeying $\gamma^{\dagger}_j = \gamma_j$, with anticommutation relations $\{ \gamma_j , \gamma_{j'} \} = \delta_{jj'}$.

The island Hamiltonian is given by 
\begin{equation}
H_{\textrm{island}} = E_C (Q-n_g)^2 - E_J \cos \Xi,
\end{equation}
where $E_C$ is the charging energy, the number operator $Q$ measures the total charge on the island (the number of Cooper pairs and occupied Majorana states), $n_g$ is the backgate parameter (assumed to be close to an integer) determined by the voltage across the capacitor, $E_J$ the Josephson energy for the coupling between the island and the bulk superconductor, where $\Xi$ is their phase difference (we will take the phase of the island to be $\Xi$, canonically conjugate to the number of Cooper pairs). The system on the island inherits a superconducting gap $\Delta_{\mathrm{sc}}$ due to proximity, which was needed for the formation of the Majoranas. We consider this energy scale to be large, so that only the Majorana bound states contribute to charge transport.

The coupling between the lead electrons and the Majorana modes on the island is given by the tunneling Hamiltonian \cite{fu2010,zazu2011} 
\begin{equation}  \label{ht}
H_t=  \sum_{j=1}^N \ \lambda_j e^{-i\Xi/2} \Psi_j^\dagger (0) \gamma_j + {\rm h.c.},
\end{equation}
where we choose the couplings $\lambda_j$ to be real and positive. This lead-Majorana tunneling gives a hybridization energy of $\Gamma_j = 2\pi \nu_0 \lambda_j^2$, where $\nu_0 = 1/\pi v_F$ is the density of states for the unperturbed leads.

In the following we will be interested in two limiting cases, where the low-energy solution of the problem simplifies \cite{beri1,emze}: for $E_J = 0$, the low-energy (i.e., for $T,V \ll E_C,\Delta_{\textrm{sc}}, \min \Gamma_j$) behavior is governed by an SO$_2$($N$) topological Kondo effect, whereas when $E_J$ is the largest energy energy scale, the topological Kondo effect has symmetry group SO$_1$($N$).

\subsection{Low-energy theory without Josephson coupling}

In the absence of Josephson coupling, i.e., with $E_J=0$, the physics at low energies ( $T,V \ll E_C,\Delta_{\textrm{sc}}, \min \Gamma_j$) is governed by virtual transitions of electrons hopping onto the dot, leading to an effective low-energy Hamiltonian $H = H_{\textrm{leads}}  + H_K^{(1)}$, where \cite{beri1}
\begin{equation}  \label{HK1}
H_K^{(1)} = \sum_{i\neq j} J_{jk}^+ \gamma_j \gamma_k \Psi^{\dagger}_k (0) \Psi_j(0) - \sum_j J_{jj}^- \Psi^{\dagger}_j (0) \Psi_j(0),
\end{equation}
for the tunneling between the leads. The (positive) coupling constants are given by $J^{\pm}_{jk} \approx \lambda_j \lambda_k / E_C$. The first term in $H_K^{(1)} $ shows a non-local quantum impurity set up by the products $ \gamma_j \gamma_k$, exchange-coupled to the spin object formed by the lead electron products $\Psi^{\dagger}_k (0) \Psi_j(0)$. The resulting entanglement gives rise to a multichannel topological Kondo effect below the energy scale defined by the Kondo temperature $T_K$; here $T_K \sim E_C e^{-1/\nu_0 J }$ when assuming isotropic $J_{jk}^+ = J$.

Including electron-electron interactions, the leads are conveniently treated using bosonization, \cite{gogolinbook} which expresses the lead Hamiltonian as
\begin{equation} \label{LL}
H_{\textrm{leads}}  = \frac{v}{2\pi} \sum_{j=1}^N \int_0^{\infty} \mathrm{d}x \left[  K (\partial_x \theta_j  )^2 + \frac{1}{K} (\partial_x \varphi_j)^2 \right],
\end{equation}
where $\theta_j$ and $\varphi_j$ are non-chiral bosonic fields with commutation relation $[ \varphi_i(x), \partial_y \theta_j(y)] = 2\pi \mathrm{i} \delta(x-y) \delta_{ij}$, $K$ is the Luttinger-liquid interaction parameter (with $K=1$ in the absence of interactions, and $K<1$ for repulsive interactions) and $v$ the interaction-renormalized Fermi velocity. The bosonized form of the electron operator is then given by $\Psi_{j,I/O} = \chi_j (2\pi a)^{-1/2} e^{\mathrm{i}(\theta_j  \mp \varphi_j )}$, where $a$ is the short-distance cut-off, and $\chi_j$ is the Klein factor (a Majorana fermion). This Majorana fermion from bosonization can be hybridized with the localized Majorana fermion $\gamma_j$ coupled to the lead, such that one simply replaces $\gamma_j \chi_j$ with a number $\pm i$ which is gauged away, see Refs.~\onlinecite{beri2,altland1}. This leads to a description of the strong-coupling fixed point in terms of the bosonic field $\mathbf{\Theta} = (\Theta_1,...,\Theta_N)$, where $\Theta_j = \theta_j(x=0)$, which is pinned by the potential
\begin{equation}
V^{(1)}[\mathbf{\Theta}] \propto - \sum_{j \neq k} \cos (\Theta_j - \Theta_k),
\end{equation}
whose minima form an $N-1$ dimensional triangular lattice. This means that in a rotated basis, the ''zero-mode'' $\check{\Theta}_0 \equiv (1/\sqrt{N}) \sum_j \Theta_j \equiv \mathbf{v}_0 \cdot \mathbf{\Theta}$, is a free field (physically, this is due to current conservation at the junction), whereas the components $\check{\Theta}_1, ... , \check{\Theta}_{N-1}$, described by vectors $\mathbf{v}_1,...,\mathbf{v}_{N-1}$ orthogonal to $\mathbf{v}_0$ (spanning the reciprocal $N-1$ dimensional triangular lattice), are fixed. Explicitly, the rotated basis is given by
\begin{eqnarray}
\check{\theta}_0 &=& \frac{1}{\sqrt{N}} \sum_{j=1}^N \theta_j, \nonumber \\
\check{\theta}_1 &=&  \frac{1}{\sqrt{2}} \theta_1 -  \frac{1}{\sqrt{2}} \theta_2, \label{rotated} \\
\check{\theta}_2 &=&  \frac{1}{\sqrt{6}} \theta_1 +  \frac{1}{\sqrt{6}} \theta_2  - \frac{2}{\sqrt{6}} \theta_3\nonumber \\ &\vdots& \nonumber \\
\check{\theta}_{N-1} &=&  \frac{1}{\sqrt{N(N-1)}} \sum_{j=1}^{N-1} \theta_j -   \frac{N-1}{\sqrt{N(N-1)}} \theta_N ,\nonumber
\end{eqnarray}
where for $N=3$ the last line should be neglected.

Hence at strong coupling we have a theory of Luttinger liquid wires (\ref{LL}) connected at a junction ($x=0$), where the field $\check{\theta}_0 (x) $ obeys Neumann (free) boundary condition (BC), whereas the orthogonal components $\check{\theta}_1(x), ... , \check{\theta}_{N-1}(x)$ obey Dirichlet (fixed) BCs. By duality, we simultaneously have that $\check{\varphi}_0 (x) $ obeys Dirichlet BC, and that the orthogonal components $\check{\varphi}_1(x), ... , \check{\varphi}_{N-1}(x)$ obey Neumann BCs (the $\check{\varphi}$ fields are obtained from the $\varphi$ fields in the same way as the $\check{\theta}$ fields from the $\theta$ fields). 

Furthermore, instanton tunneling of the pinned fields at strong coupling yields a leading irrelevant operator with scaling dimension \cite{beri2,altland1} $\Delta_{LIO} = 2K(N-1)/N$, determining the finite-temperature scaling of the non-local conductance.

\subsection{Low-energy theory with strong Josephson coupling}

Another type of low-energy topological Kondo effect is obtained in the limit of strong Josephson coupling, more specifically when $\max \Gamma_j \ll \sqrt{8E_C E_J} \alt E_J$; see Ref.~\onlinecite{emze}. The low-energy theory that emerges in this parameter regime is given by  $H = H_{\textrm{leads}}  + H_A + H_K^{(2)} $, where
\begin{eqnarray}
H_A &=& - \sum_j \lambda_j \gamma_j \Psi^{\dagger}_j(0) + \mathrm{h.c.},\label{RAR}  \\
 H_K^{(2)} &=& \sum_{j\neq k} J_{jk} \gamma_j \gamma_k  (\Psi^{\dagger}_k (0) + \Psi_k (0) ) (\Psi^{\dagger}_j (0) + \Psi_j (0) ), \nonumber \\
\end{eqnarray}
where $\lambda_j$ is the Majorana tunneling coupling in (\ref{ht}) and $J_{jk} \approx \lambda_j \lambda_k / E_J$. Here $H_A$ corresponds to the usual single-lead resonant Andreev reflection processes, while the exchange term $H_K^{(2)}$ contains both the same processes as in (\ref{HK1}) as well as crossed Andreev reflection processes.

Performing the same bosonization procedure as above for $H_K^{(1)}$ now leads to a strong-coupling pinning potential \cite{emze}
\begin{equation}
V^{(2)}[\mathbf{\Theta}] \propto - \sum_j \sqrt{\Gamma_j} \sin \Theta_j - \sqrt{T_K} \sum_{j \neq k} \cos \Theta_j \cos \Theta_k,
\end{equation}
for the $\mathbf{\Theta}$ field. This implies a manifold of strong-coupling fixed points, tuned by the $N$ parameters $\delta_j \equiv \sqrt{\Gamma_j / T_K}$, where the minima of the potential $V^{(2)}[\mathbf{\Theta}]$ form an $N$ dimensional generalization of the body-centered cubic lattice for $\Gamma_j \ll T_K$, with the center-point being shifted as a function of the $\delta_j$ parameters. Here the Kondo temperature $T_K$ defines the energy scale below which the Kondo effect develops, given by $T_K \approx \sqrt{8E_J E_C} e^{-E_J / (N-2)\Gamma}$ for isotropic $\Gamma_j = \Gamma$.

Hence in the regime of strong Josephson coupling, the strong-coupling theory is that of Luttinger liquid wires connected at a junction where all the fields $\check{\theta}_0 (x), \check{\theta}_1(x), ... , \check{\theta}_{N-1}(x) $ have Dirichlet BCs, and all the dual fields $\check{\varphi}_0 (x), \check{\varphi}_1(x), ... , \check{\varphi}_{N-1}(x) $ have Neumann BCs.

The finite-temperature behavior is governed by a leading irrelevant operator with scaling dimension 
\begin{equation}\label{scaldim2}
\Delta_{LIO} ={\rm min}\left\{ 2, \frac12 \sum_{j=1}^N   
\left[ 1- \frac{2}{\pi} \sin^{-1}  \left(
\frac{\delta_j}{2(N-1)}\right) \right]^2\right\} ,
\end{equation}
arising from instanton tunneling of the fields between adjacent potential minima.

\section{Local density of states}  \label{chap:ldos}

The local density of states $\rho_i$ available for electron tunneling into the $i$th lead is given by

\begin{eqnarray}  \label{ldos}
\rho_i(x,\omega) &=& -\frac{1}{\pi} \mathrm{Im} \, G_i^R(x,\omega) \nonumber \\
&=&\frac{ 1}{\pi} \, \mathrm{Re} \int_0^{\infty} \mathrm{d}  t  \, e^{\mathrm{i} \omega t} \langle \Psi_i (x,t)  \Psi_i^{\dagger} (x,0) \rangle,
\end{eqnarray}
where $G_i^R(x,\omega)$ is the equal-position retarded Green's function for the electrons in the $i$th lead. The local density of states $\rho_i$ is directly measurable using scanning tunneling microscopy, as the differential tunneling conductance $G_i(x,V)$ at position $x$ in lead $i$ is directly proportional to this quantity as a function of applied voltage $V$, i.e. $G_i(x,V) \propto \rho_i(x,\omega = eV)$.

\begin{figure}[h]
\centering
\includegraphics[width=8cm]{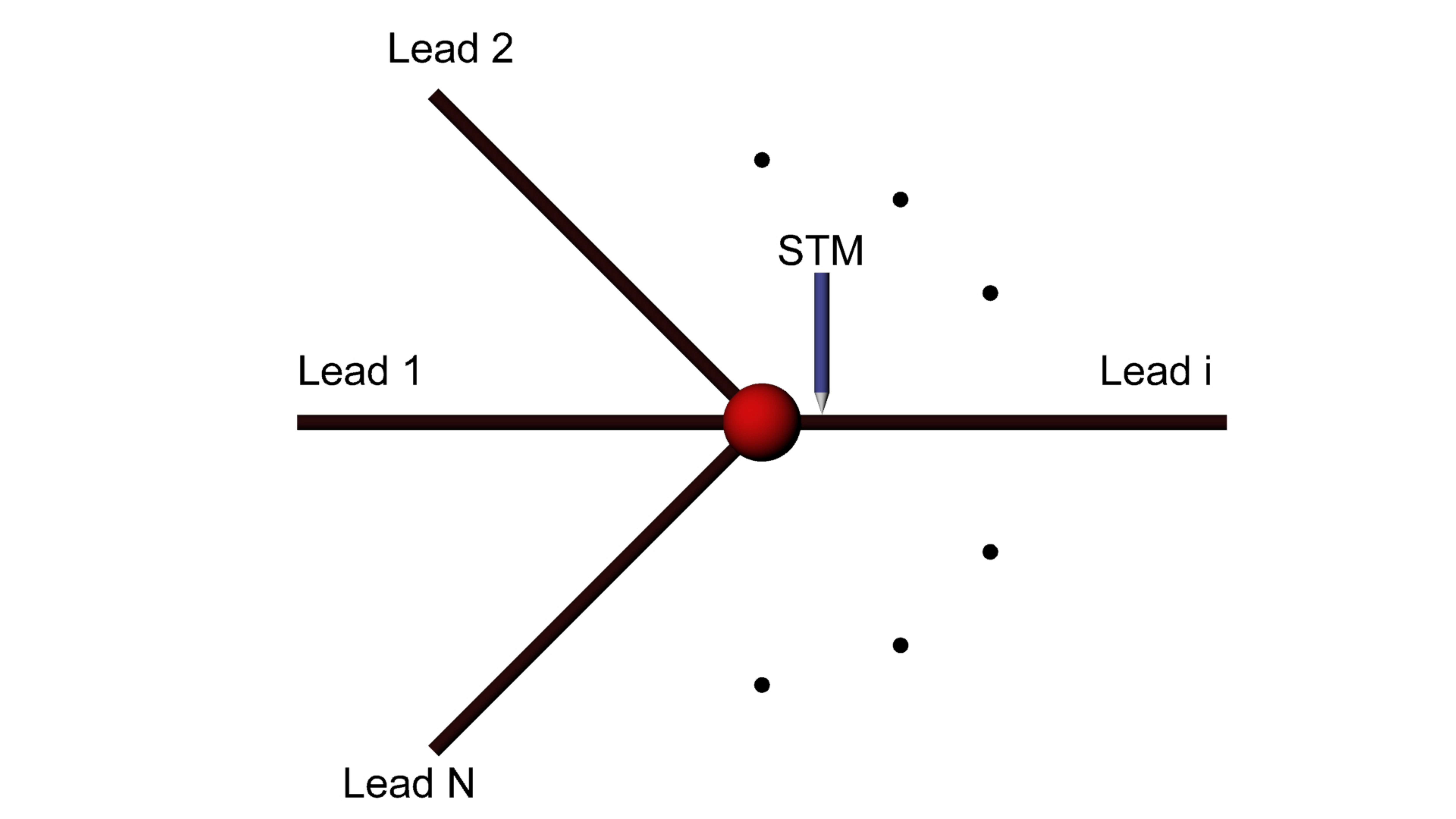}
\caption{\label{fig:2} 
(Color online) The topological Kondo problem at low energy is equivalent to an $N$-wire junction with a splitting matrix $\mathbb{M}$ describing the boundary condition at the junction. With an STM tip the LDOS $\rho_i(x,\omega)$ of  wire $i$ is probed. }
\end{figure}

We shall here be concerned with the low-energy behavior of the LDOS, where temperature $T$ and energy $\omega$ are well below the Kondo temperature $T_K$ of the system. With the $N$ wires effectively connected at a single junction with a boundary condition due to the topological Kondo effect, see Fig.~\ref{fig:2}, the problem of finding the LDOS is analogous to that for a junction of several Luttinger liquid wires. \cite{NFLL,cte,oshikawa,hou,adrs,hou2}

\subsection{Electron Green's function}  \label{Green}
The zero-temperature, equal-position Green's function $\langle \Psi_i (x,t)  \Psi_i^{\dagger} (x,0) \rangle$ for wire $i$ in the $N$-wire junction system can be calculated following Agarwal {\em et al.}~in Ref.~\onlinecite{adrs}. This amounts to finding the current-splitting matrix $\mathbb{M}$ for the junction, which relates the incoming $j_{i,I}$ and outgoing $j_{i,O}$ currents at the junction through $ j_{i,O} = \sum_j \mathbb{M}_{ij}  j_{j,I}$. In terms of the chiral bosonic fields $\phi_{i,I} =  \theta_i - \varphi_i $ and $\phi_{i,O} =   \theta_i + \varphi_i$, such that the electron field is expressed as $\Psi_{j,I/O} \propto e^{\mathrm{i}\phi_{j,I/O}}$, the $\mathbb{M}$ matrix is equivalent to the boundary condition 
\begin{equation}
\phi_{i,O} = \sum_j \mathbb{M}_{ij} \phi_{j,I}.
\end{equation}
With a Bogoliubov transformation 
\begin{equation}
\phi_{j,O/I} = [ (1+K)\tilde{\phi}_{j,O/I} + (1-K)\tilde{\phi}_{j,I/O} ] / (2\sqrt{K}),
\end{equation}
one obtains the free outgoing/incoming fields $\tilde{\phi}_{j,O/I} $ with commutation relations 
\begin{equation}
[\tilde{\phi}_{j,O/I} (x,t) , \tilde{\phi}_{j,O/I} (x',t)] = \pm \mathrm{i} \pi\, \mathrm{sgn}(x-x').
\end{equation}
Their splitting matrix $\tilde{\mathbb{M}}$, which relates $\tilde{\phi}_{i,O}(x) = \sum_j \tilde{\mathbb{M}}_{ij} \tilde{\phi}_{j,I}(-x) $ in the ''unfolded picture'' (where $x$ is extended to the entire real line) is given by 
\begin{equation}
\tilde{\mathbb{M}} = [ (1+K) \mathbb{M} + (1-K)\mathbb{I} ]  [  (1+K) \mathbb{I} + (1-K)\mathbb{M}  ]^{-1},
\end{equation}
where $\mathbb{I}$ is the identity matrix. 

The Green's function now follows from
\begin{eqnarray}
&& \langle \Psi_i (x,t)  \Psi_i^{\dagger} (x,0) \rangle  = \langle \Psi_{i,I} (x,t)  \Psi_{i,I}^{\dagger} (x,0) \rangle \nonumber \\
&& \quad +\  \langle \Psi_{i,O} (x,t)  \Psi_{i,O}^{\dagger} (x,0) \rangle + e^{\mathrm{i} 2k_F x} \langle \Psi_{i,O} (x,t)  \Psi_{i,I}^{\dagger} (x,0) \rangle \nonumber \\
&& \quad +  \ e^{-\mathrm{i} 2k_F x} \langle \Psi_{i,I} (x,t)  \Psi_{i,O}^{\dagger} (x,0) \rangle ,
\end{eqnarray}
where the two oscillatory terms vanish for lead lengths $L \to \infty$ in the cases we are interested in, since the corresponding Green's functions contain an $L$ dependence $ \sim L^{(\tilde{\mathbb{M}}_{ii} -1)K}$. The remaining terms are given by
\begin{eqnarray}
&&\langle \Psi_{i,O} (x,t)  \Psi_{i,O}^{\dagger} (x,0) \rangle  = \frac{1}{2\pi a} \langle e^{\mathrm{i} \phi_{ i,O} (x,t) }  e^{-\mathrm{i} \phi_{ i,O} (x,0) }    \rangle \nonumber \\
&& \quad = \frac{1}{2\pi a} \langle e^{\mathrm{i} [ (1+K)\tilde{\phi}_{i,O}(x,t) + (1-K) \tilde{\phi}_{i,I}(x,t) ] / (2\sqrt{K}) } \nonumber \\
&& \quad \qquad  \times e^{-\mathrm{i} [ (1+K)\tilde{\phi}_{i,O}(x,0) + (1-K) \tilde{\phi}_{i,I}(x,0) ] / (2\sqrt{K}) }   \rangle \nonumber \\
&& \quad = \frac{1}{2\pi a} \langle e^{\mathrm{i} [ (1+K)  \sum_j \tilde{\mathbb{M}}_{ij} \tilde{\phi}_{j,I}(-x,t) } e^{\mathrm{i} (1-K) \tilde{\phi}_{i,I}(x,t) ] / (2\sqrt{K}) } \nonumber \\
&& \quad \quad \ \  \times e^{-\mathrm{i} [ (1+K) \sum_j \tilde{\mathbb{M}}_{ij} \tilde{\phi}_{j,I}(-x,0) } e^{-\mathrm{i}  (1-K) \tilde{\phi}_{i,I}(x,0) ] / (2\sqrt{K}) }   \rangle. \nonumber \\ \label{gf2}
\end{eqnarray}
With the relation $\langle e^{\mathrm{i} \alpha_1 \phi(z_1) } \cdots e^{\mathrm{i} \alpha_n \phi(z_n) } \rangle = \prod_{i<j} (z_i - z_j ) ^{\alpha_i \alpha_j}$ for the expectation value of a product of vertex operators with complex coordinates $z = x + \mathrm{i} \tau$, \cite{cft} one arrives at $\langle \Psi_{i,O} (x,t)  \Psi_{i,O}^{\dagger} (x,0) \rangle = \langle \Psi_{i,I} (x,t)  \Psi_{i,I}^{\dagger} (x,0) \rangle = \langle \Psi_i (x,t)  \Psi_i^{\dagger} (x,0) \rangle  / 2$, with \cite{adrs}
\begin{eqnarray}  \label{gf}
&& \langle \Psi_i (x,t)  \Psi_i^{\dagger} (x,0) \rangle =  \nonumber \\
&&\qquad \qquad \qquad = \frac{1}{2\pi a} \left[ \frac{\mathrm{i} a }{-vt + \mathrm{i} a} \right]^{(K+1/K)/2} \\
&& \qquad\qquad \qquad \qquad  \times \left[ \frac{-a^2 -4x^2}{(-vt + \mathrm{i} a)^2 - 4x^2} \right]^{\tilde{\mathbb{M}}_{ii} (1/K-K)/4}. \nonumber
\end{eqnarray}

Now, close to the junction, where we can put $x\to 0$, as well as far from the junction, where $x\to \infty$, the expressions allow us to compute the LDOS. When $x\to 0$, we have
\begin{eqnarray}  \label{gf0}
&& \langle \Psi_i (0,t)  \Psi_i^{\dagger} (0,0) \rangle \nonumber \\
&& \qquad \qquad = \frac{1}{2\pi a} \left[ \frac{\mathrm{i} a }{-vt + \mathrm{i} a} \right]^{\{ (1-\tilde{\mathbb{M}}_{ii} ) K+(1+\tilde{\mathbb{M}}_{ii} ) /K \}/2}, \qquad  \label{gf3}
\end{eqnarray}
which means that the (chiral) boundary field $\Psi_i (0,t)$ has scaling dimension 
\begin{equation}  \label{delta}
\Delta_i = \{ (1-\tilde{\mathbb{M}}_{ii} ) K+(1+\tilde{\mathbb{M}}_{ii} ) /K \}/2,
\end{equation}
i.e. $\langle \Psi_i (0,\tau)  \Psi_i^{\dagger} (0,0) \rangle \sim \tau^{-\Delta_i}$ for imaginary time $\tau \gg a/v$. 

Similarly, far away from the junction, where $x \to \infty$, one has 
\begin{eqnarray}  \label{gfi}
&& \langle \Psi_i (x,t)  \Psi_i^{\dagger} (x,0) \rangle  =\frac{1}{\pi a} \left[ \frac{\mathrm{i} a }{-vt + \mathrm{i} a} \right]^{(K+1/K)/2}  ,
\end{eqnarray}
implying the usual scaling exponent $\Delta_i = (K+1/K)/2$ for bulk (non-chiral) electrons.

\subsection{The local density of states} \label{sec:ldos}

Far away from the junction, putting Eq.~(\ref{gfi}) for the Green's function ($x\to \infty$) into the expression (\ref{ldos}) for the LDOS, we arrive at \cite{adrs,bruus}
\begin{equation}  \label{xinf}
\rho_i(x\to \infty,\omega) = \frac{1}{a\pi \Gamma(\Delta_i)} \left( \frac{a}{v} \right)^{\Delta_i} \omega^{\Delta_i -1} e^{-a \omega / v } H(\omega),
\end{equation}
where $\Gamma$ is the gamma function and $H$ is the Heaviside step function, and with the above scaling dimension $\Delta_i = (K+1/K)/2$. For non-interacting electrons in the leads this reduces to $\rho_i(x,\omega) = 1/(\pi v) \equiv \nu_0$, i.e. the density of states $\nu_0$ for a bulk spinless quantum wire, as expected.

Considering positive energies $\omega \ll v/a$, we will neglect the factor $e^{-a \omega / v } H(\omega)$ in the discussion below.

An analytical expression can also be obtained for the limit $2x\omega /v \gg 1$, resulting in \cite{LL}
\begin{eqnarray} \label{xomega}
\rho_i(x,\omega) &=& \frac{1}{\pi v \Gamma((K+1/K)/2)  } \left( \frac{a\, \omega}{v} \right)^{(K+1/K)/2 -1} \nonumber \\  \label{x}
&& \  + \frac{2^{2-(K+1/K)/2} \cos (2x\omega /v + \delta) }{\pi v \Gamma ( \tilde{\mathbb{M}}_{ii}(1/K-K)/4)}  \\
&&  \quad \times  \left( \frac{a\, \omega }{v} \right)^{ [\tilde{\mathbb{M}}_{ii}(1/K-K)/4] -1} \nonumber \\
&& \quad \times  \left( \frac{a}{x} \right)^{(3K + 1/K)(1+\tilde{\mathbb{M}}_{ii})/8 + (K + 3/K)(1-\tilde{\mathbb{M}}_{ii})/8 } \nonumber
\end{eqnarray}
where $\delta \equiv \mathrm{Arg} (\mathrm{i}^{(K + 3/K)(1+\tilde{\mathbb{M}}_{ii})/8 + (3K + 1/K)(1-\tilde{\mathbb{M}}_{ii})/8} ) $. Note that for fixed $\omega$ the second term vanishes as $x\to \infty$, reducing the expression (\ref{xomega}) to that in Eq.~(\ref{xinf}).

Finally and most importantly, namely close to the junction, putting the expression (\ref{gf0}) for the Green's function of the chiral boundary field at the junction ($x=0$) into the expression (\ref{ldos}) for the LDOS, we arrive at \cite{adrs}
\begin{equation}  \label{LDOS}
\rho_i(0,\omega) = \frac{1}{a2\pi \Gamma(\Delta_i)} \left( \frac{a}{v} \right)^{\Delta_i} \omega^{\Delta_i -1},
\end{equation}
with $\Delta_i$ now given by Eq.~(\ref{delta}). This behavior occurs within a distance of the order of $x < v/(2\omega)$ from the junction.

In order to proceed, we must now see what values for the $\tilde{\mathbb{M}}$ matrix the different boundary conditions in the topological Kondo effect correspond to.  

\subsection{Local density of states for topological Kondo systems} \label{sec:ldos2}

\subsubsection{Strong Josephson coupling}

The simplest case is for strong Josephson coupling, where all the $\check{\theta}_j$ fields have Dirichlet, and all the $\check{\varphi}_j$ fields have Neumann BCs, at all strong-coupling fixed points. The electron operator $\Psi_{j,I/O} \propto e^{\mathrm{i}(\theta_j \mp \varphi_j )}$ at the junction at $x=0$ is then given by
\begin{equation}
\Psi_{j,O}(0) \propto e^{\mathrm{i}[ \theta_j(0) + \varphi_j(0) ]} = e^{\mathrm{i}c_i} e^{  \mathrm{i}\varphi_j(0)},
\end{equation}
where $c_i$, a constant depending on the potential minimum the $\theta_j(0)$ field is trapped in, can be gauged away. Hence $\Psi_{j,O}(0) = \Psi_{j,I}^{\dagger}(0) $, meaning that $\tilde{\phi}_{i,O}(0) = - \tilde{\phi}_{i,I}(0)  $, i.e. the $\tilde{\mathbb{M}}$ matrix is that for perfect Andreev reflection in each lead separately, namely
\begin{equation}
\tilde{\mathbb{M}} = \left(
\begin{array}{cccc}
-1& 0& \ldots &0  \\
0 & -1& \ldots & 0\\
\vdots & \vdots & \ddots & \vdots \\
0& 0 & \ldots & -1
\end{array} \right),
\end{equation}
such that $\tilde{\mathbb{M}}_{ii} = -1$ for all $i$. 

Let us now consider the electron Green's function (\ref{gf}) close to the junction, i.e. letting $x \to 0$. With $\tilde{\mathbb{M}}_{ii} = -1$,
\begin{eqnarray}  
&& \langle \Psi_i (x,t)  \Psi_i^{\dagger} (x,0) \rangle = \frac{1}{2\pi a} \left[ \frac{\mathrm{i} a }{-vt + \mathrm{i} a} \right]^{K} 
\end{eqnarray}
implying a scaling dimension (\ref{delta}) equal to $\Delta_i = K$. The lead LDOS at the junction therefore behaves as 
\begin{equation}  \label{LDOSK}
\rho_i(0,\omega) \sim \omega^{K-1}.
\end{equation}
Hence the LDOS has exactly the same behavior as for a single-wire perfect Andreev reflection, \cite{Fidkowski} meaning that tunneling spectroscopy follows the same power law for all fixed points appearing, i.e. there is no difference between the Kondo fixed point manifold and the resonant Andreev reflection fixed point.

For non-interacting lead electrons, i.e. with $K=1$, Eq.~(\ref{LDOS}) results in
\begin{equation}  \label{K1}
\rho_i(0,\omega) = \frac{1}{2\pi v} = \frac{\nu_0}{2}, \qquad \qquad K=1,
\end{equation}
such that the electron density of states at the junction is half of that for bulk spinless electrons. 

This can be confirmed by the exact solution for a Majorana fermion coupled to a quantum wire. Decomposing the lead electron into two Majorana fermions $\eta$ and $\zeta$, such that $ \Psi_j(x) = [ \eta_j(x) + \mathrm{i} \zeta_j(x) ] /\sqrt{2}$, the Majorana tunneling term (\ref{RAR}) reads $H_A \propto \sum_j  \sqrt{\Gamma_j} \gamma_j \zeta_j(0)$. Hence at the resonant Andreev reflection fixed point ($\Gamma_j \to \infty$), the $\zeta_j$ Majorana is hybridized with the $\gamma_j$ Majorana within a ''screening cloud'' of size \cite{ag} $\xi_M \sim v/\Gamma_j$. In particular, the $x=0$ Matsubara Green's function $G_{\zeta_j}$ for the $\zeta_j$ Majorana is given by \cite{lehur}
\begin{equation}
G_{\zeta_j} (0,\mathrm{i}\omega_n) = \frac{-\mathrm{i} \,\mathrm{sgn}(\omega_n)}{2 v}  \frac{\mathrm{i} \omega_n}{\mathrm{i} \omega_n + \mathrm{i} \Gamma_j \mathrm{sgn} (\omega_n) }.
\end{equation}
Hence the $\zeta_j$ contribution $\propto \mathrm{Im} \, G_{\zeta_j} (0,\mathrm{i}\omega_n \to \omega)$ to the LDOS vanishes as $\Gamma_j \to \infty$.

Therefore, at $x\ll \xi_M$, only the $\eta_j$ Majorana contributes to the LDOS of the lead electron, which thus is half the bulk value, i.e. $\rho_j(0,\omega) = \nu_0 /2$.

\subsubsection{Without Josephson coupling}

For the topological Kondo model without Josephson coupling, i.e. the SO$_2$(N) model of B\'eri and Cooper, \cite{beri1} the fields $\check{\theta}_0 (x), \check{\varphi}_1(x), ... , \check{\varphi}_{N-1}(x) $ have Neumann BCs, and the fields $\check{\varphi}_0(x),\check{\theta}_1(x), ... , \check{\theta}_{N-1}(x)$ Dirichlet BCs at the strong-coupling fixed point. 

The original fields in terms of the rotated ones in Eq.~(\ref{rotated}) are given by
\begin{eqnarray}
\theta_1 &=&  \frac{1}{\sqrt{N}} \check{\theta}_0 +  \frac{1}{\sqrt{2}}  \check{\theta}_1 + \frac{1}{\sqrt{6}}  \check{\theta}_2 + ... +  \frac{1}{\sqrt{N(N-1)}}  \check{\theta}_{N-1}, \nonumber \\
\theta_2 &=&  \frac{1}{\sqrt{N}} \check{\theta}_0 -  \frac{1}{\sqrt{2}}  \check{\theta}_1 + \frac{1}{\sqrt{6}}  \check{\theta}_2 + ... +  \frac{1}{\sqrt{N(N-1)}}  \check{\theta}_{N-1}, \nonumber \\
\theta_3 &=&  \frac{1}{\sqrt{N}} \check{\theta}_0 - \frac{2}{\sqrt{6}}  \check{\theta}_2 + ... +  \frac{1}{\sqrt{N(N-1)}}  \check{\theta}_{N-1}, \nonumber \\
&\vdots&  \\
\theta_{N} &=&  \frac{1}{\sqrt{N}} \check{\theta}_0  - \frac{N-1}{\sqrt{N(N-1)}}   \check{\theta}_{N-1}    ,\nonumber
\end{eqnarray}
where for $N=3$ the terms after the dots should be neglected. The change of basis between $ \varphi_j$ and $ \check{\varphi}_j$ is the same.

Hence, the electron operator $\Psi_{j,I/O} \propto e^{\mathrm{i}(\theta_j \mp \varphi_j  )}$ at the junction at $x=0$ is then given by, for simplicity considering lead $j=1$,
\begin{eqnarray}
\Psi_{1,O}(0) & \propto &  e^{\mathrm{i}[\varphi_1(0)+ \theta_1(0)]} =  e^{\mathrm{i}[\frac{1}{\sqrt{N}} \check{\varphi}_0(0) + ... + \frac{1}{\sqrt{N} }\check{\theta}_0(0) + ...]} \nonumber \\
&=& e^{\mathrm{i}c_1} e^{\mathrm{i}[\frac{1}{\sqrt{N} }\check{\theta}_0(0) +  \frac{1}{\sqrt{2} }\check{\varphi}_1(0) + ... + \frac{1}{\sqrt{N(N-1)} }\check{\varphi}_{N-1}(0)  ]}, \nonumber \\ \label{psi1}
\end{eqnarray}
with $c_1 $ a constant, depending on the pinning value of the fields with Dirichlet BCs, which we gauge away.

From Eqs.~(\ref{gf2})-(\ref{gf3}) it follows that the term $\check{\theta}_0/\sqrt{N}$ in the exponent in Eq.~(\ref{psi1}) contributes a term $1/(NK)$, and each term $\check{\varphi}_n/\sqrt{n(n+1)}$ contributes a term $K /  [n(n+1)]$, in the exponent of $\langle \Psi_1 (0,t)  \Psi_1^{\dagger} (0,0) \rangle$, which gives
\begin{eqnarray}  
&& \langle \Psi_1 (0,t)  \Psi_1^{\dagger} (0,0) \rangle   \nonumber \\
&& \qquad \qquad = \frac{1}{\pi a} \left[ \frac{\mathrm{i} a }{-vt + \mathrm{i} a} \right]^{  \frac{1}{NK}  + \sum_{k=1}^{N-1} \frac{1}{k(k+1)} K    } \qquad 
\end{eqnarray}
(see also the Appendix for a derivation of the $\tilde{\mathbb{M}}$ matrix). Hence we have the scaling exponent
\begin{equation}  \label{scaldim}
\Delta_i = \frac{1}{NK} + \frac{N-1}{N} K.
\end{equation}
For $x \ll v/(2\omega)$, the lead LDOS therefore goes as
\begin{equation} \label{ldos1}
\rho_i(x \to 0,\omega) \sim \omega^{\frac{1}{NK} + \frac{N-1}{N} K-1}.
\end{equation}
Thus, for $\frac{1}{N-1} < K <1$, we have a diverging LDOS at zero energy in the vicinity of the junction. For non-interacting lead electrons, $K=1$, we get $\Delta_i =1$, again giving the result $\rho_i(0,\omega) = \frac{\nu_0}{2}$ according to Eq.~(\ref{LDOS}).

Note also, that in the $2x\omega /v \gg 1$ limit, there is an unusual exponent in the $x$ dependence of the subleading oscillatory term in Eq.~(\ref{x}), which has an envelope decaying as $\sim x^{- 3/(4K) - (K-1/K)/(2N) }$ as a function of distance $x$ from the junction, and diverging as $\sim \omega^{(1-2/N)(K-1/K)/4 -1}$ as function of energy.

\section{Discussion}

In this work, we have investigated the tunneling spectroscopy of topological Kondo systems, providing a route complementary to transport measurements in the search for experimental signatures of the predicted non-Fermi liquid behavior.

We have found that for the minimal topological Kondo setup of B\'eri and Cooper, \cite{beri1} with a strong-coupling SO$_2$($N$) Kondo fixed point, the LDOS of the effectively spinless electrons in lead $i$ in the immediate neighborhood of the junction (meaning that the distance $x$ from the junction is less than $v/(2\omega)$) follows the power law in Eq.~(\ref{ldos1}), i.e. it goes as $ \sim \omega^{\frac{1}{NK} + \frac{N-1}{N} K-1}$ as a function of energy $\omega$. For non-interacting leads, $K=1$, the LDOS close to the junction is a constant, equal to half the bulk value, i.e. $1/(2\pi v)$. However, for interacting lead electrons, $K<1$, the scaling dimension (\ref{scaldim}) controlling the LDOS and hence the tunneling conductance of an STM tip probing lead $i$, depends on the number $N$ of leads. An experimental signature of the topological Kondo fixed point is therefore obtained by, using gate voltages, changing the number $N$ of leads coupling to the Majoranas on the island, and then observing how the scaling exponent of the tunneling conductance in lead $i$ changes.

In the topological Kondo system with a strong Josephson coupling, realizing an SO$_1$($N$) topological Kondo fixed point together with a resonant Andreev reflection fixed point and a continuous manifold of fixed points where Kondo and resonant Andreev reflection processes coexist, \cite{emze} we find that the LDOS of the lead electrons close to the junction instead follows the power law $\sim \omega^{K-1}$ as a function of energy, also with the constant value $1/(2\pi v)$ for $K=1$. Hence in the strong Josephson coupling case, an STM experiment cannot distinguish the Kondo fixed point, or the coexistence manifold, from the pure resonant Andreev reflection fixed point.

The only trace of the topological Kondo physics in the LDOS in the SO$_1$($N$) case would come from the corrections due to the leading irrelevant operators at the fixed points. With scaling dimension $\Delta_{\textrm{LIO}} > 1$, given by Eq.~(\ref{scaldim2}), these operators contribute terms $\sim \omega^{\Delta_{\textrm{LIO}}  - 1}$ to the LDOS at $x \to 0$. Hence in these subleading corrections there is a difference between the resonant Andreev reflection fixed point where $\Delta_{\textrm{LIO}} =2$ and in the Kondo fixed point manifold, where $1 < \Delta_{\textrm{LIO}} \leq 3/2$ ($1 < \Delta_{\textrm{LIO}} \leq 2$) for $N=3$ ($N >3$). However, any repulsive interaction among the lead electrons renders the LDOS (\ref{LDOSK}) divergent at zero energy, obscuring the subleading corrections which vanish as $\omega \to 0$.

In summary, we have provided analytical expressions for the LDOS of the leads in Majorana devices hosting the topological Kondo effect. This provides a clear signature, complementary to previously proposed transport measurements, to look for in experiments.

\section*{Acknowledgments}

We thank D. Giuliano, H. Johannesson and A. Zazunov for interesting discussions and comments.

Financial support from the SFB TR-12 and SPP 1666 of the Deutsche Forschungsgemeinschaft (E.E. and R.E.) as well as European Commission, European Social Fund and Regione Calabria (A.N.) is acknowledged.

\section*{Appendix A: Splitting matrix for topological Kondo}

Let us here compute the $\tilde{\mathbb{M}}$ matrix for the topological Kondo effect of B\'eri and Cooper. \cite{beri1}

First, note that the non-chiral and chiral bosonic fields (see Sec.~\ref{Green}) are related by

\noindent 
\begin{eqnarray}
\tilde{\varphi}_{i}(x)&=&\frac{1}{\sqrt{K}}\varphi_{i}(x)=\left(\tilde{\phi}_{O,i}-\tilde{\phi}_{I,i}\right)/2\nonumber \\ &=&\frac{1}{\sqrt{K}}\left(\phi_{O,i}-\phi_{I,i}\right)/2 \label{eq:theta},
\end{eqnarray}

\noindent 
\begin{eqnarray}
\tilde{\theta}_{i}(x)&=&\sqrt{K}\theta_{i}(x) = \left(\tilde{\phi}_{I,i}+\tilde{\phi}_{O,i}\right)/2 \nonumber \\
&=&\sqrt{K}\left(\phi_{I,i}+\phi_{O,i}\right)/2 . \label{eq:phi}
\end{eqnarray}

\begin{widetext}

The topological Kondo BC, i.e.~the fields $\check{\theta}_0 (x), \check{\varphi}_1(x), ... , \check{\varphi}_{N-1}(x) $ having Neumann BCs and the fields $\check{\varphi}_0(x),\check{\theta}_1(x), ... , \check{\theta}_{N-1}(x)$ Dirichlet BCs, means that we pin the following vector (cf.~Refs.~\onlinecite{oshikawa,hou})

\begin{equation}
 \left(\begin{array}{c}
\frac{1}{\sqrt{N}}[\tilde{\varphi}_{1}(x=0) +\tilde{\varphi}_{2}(x=0) +\tilde{\varphi}_{3}(x=0)+\ldots+\tilde{\varphi}_{N}(x=0)]\\
\frac{1}{\sqrt{2}}[\tilde{\theta}_{1}(x=0)-\tilde{\theta}_{2}(x=0)]\\
\frac{1}{\sqrt{6}}[\tilde{\theta}_{1}(x=0)+\tilde{\theta}_{2}(x=0)-2\tilde{\theta}_{3}(x=0)]\\
\frac{1}{\sqrt{12}}[\tilde{\theta}_{1}(x=0)+\tilde{\theta}_{2}(x=0)+\tilde{\theta}_{3}(x=0)-3\tilde{\theta}_{4}(x=0)]\\
\vdots\\
\frac{1}{\sqrt{(N-1)N}}[ \tilde{\theta}_{1}(x=0)+\tilde{\theta}_{2}(x=0)+\tilde{\theta}_{3}(x=0)+\tilde{\theta}_{4}(x=0)+\cdots-(N-1)\tilde{\theta}_{N}(x=0)]
\end{array}\right)  = \overrightarrow{0}
\end{equation}
to a value that we set to be the null vector $\vec{0}$. With the notation $\tilde{\Theta}_j \equiv \tilde{\theta}_j(x=0)$ and $\tilde{\Phi}_j = \tilde{\varphi}_j(x=0)$, we write this as

\begin{equation}
 \left(\begin{array}{c}
\frac{1}{\sqrt{N}}(\tilde{\Phi}_{1}+\tilde{\Phi}_{2}+\tilde{\Phi}_{3}+\ldots+\tilde{\Phi}_{N})\\
\frac{1}{\sqrt{2}}(\tilde{\Theta}_{1}-\tilde{\Theta}_{2})\\
\frac{1}{\sqrt{6}}(\tilde{\Theta}_{1}+\tilde{\Theta}_{2}-2\tilde{\Theta}_{3})\\
\frac{1}{\sqrt{12}}(\tilde{\Theta}_{1}+\tilde{\Theta}_{2}+\tilde{\Theta}_{3}-3\tilde{\Theta}_{4})\\
\vdots\\
\frac{1}{\sqrt{(N-1)N}}(\tilde{\Theta}_{1}+\tilde{\Theta}_{2}+\tilde{\Theta}_{3}+\tilde{\Theta}_{4}+\cdots-(N-1)\tilde{\Theta}_{N})
\end{array}\right)=\overrightarrow{0}.
\end{equation}

From Eqs.~(\ref{eq:phi}) and (\ref{eq:theta}) we have

\begin{equation}
\left(\begin{array}{c}
(\tilde{\Phi}_{O,1}-\tilde{\Phi}_{I,1}+\tilde{\Phi}_{O,2}-\tilde{\Phi}_{I,2}+\tilde{\Phi}_{O,3}-\tilde{\Phi}_{I,3}+\ldots+\tilde{\Phi}_{O,N}-\tilde{\Phi}_{I,N})\\
(\tilde{\Phi}_{O,1}+\tilde{\Phi}_{I,1}-\tilde{\Phi}_{O,2}-\tilde{\Phi}_{I,2})\\
(\tilde{\Phi}_{O,1}+\tilde{\Phi}_{I,1}+\tilde{\Phi}_{O,2}+\tilde{\Phi}_{I,2}-2\tilde{\Phi}_{O,3}-2\tilde{\Phi}_{I,3})\\
(\tilde{\Phi}_{O,1}+\tilde{\Phi}_{I,1}+\tilde{\Phi}_{O,2}+\tilde{\Phi}_{I,2}+\tilde{\Phi}_{O,3}+\tilde{\Phi}_{I,3}-3\tilde{\Phi}_{O,4}-3\tilde{\Phi}_{I,4})\\
\vdots\\
(\tilde{\Phi}_{O,1}+\tilde{\Phi}_{I,1}+\tilde{\Phi}_{O,2}+\tilde{\Phi}_{I,2}+\tilde{\Phi}_{O,3}+\tilde{\Phi}_{I,3}+\tilde{\Phi}_{O,4}+\tilde{\Phi}_{I,4}+\cdots-(N-1)\tilde{\Phi}_{O,N}-(N-1)\tilde{\Phi}_{I,N})
\end{array}\right)=\overrightarrow{0},
\end{equation}

\vspace{0.5cm}

where $\tilde{\Phi}_{O/I,j} = \tilde{\phi}_{O/I,j}(x=0)$. Hence

\vspace{0.5cm}

\begin{eqnarray}
\left(\begin{array}{c}
\tilde{\Phi}_{O,1}+\tilde{\Phi}_{O,2}+\tilde{\Phi}_{O,3}+\tilde{\Phi}_{O,N}\\
\tilde{\Phi}_{O,1}-\tilde{\Phi}_{O,2}\\
\tilde{\Phi}_{O,1}+\tilde{\Phi}_{O,2}-2\tilde{\Phi}_{O,3}\\
\tilde{\Phi}_{O,1}+\tilde{\Phi}_{O,2}+\tilde{\Phi}_{O,3}-3\tilde{\Phi}_{O,4}\\
\vdots\\
\tilde{\Phi}_{O,1}+\tilde{\Phi}_{O,2}+\tilde{\Phi}_{O,3}+\tilde{\Phi}_{O,4}+\cdots-(N-1)\tilde{\Phi}_{O,N}
\end{array}\right) & = & \left(\begin{array}{c}
\tilde{\Phi}_{I,1}+\tilde{\Phi}_{I,2}+\tilde{\Phi}_{I,3}+\tilde{\Phi}_{I,N}\\
-\tilde{\Phi}_{I,1}+\tilde{\Phi}_{I,2}\\
-\tilde{\Phi}_{I,1}-\tilde{\Phi}_{I,2}+2\tilde{\Phi}_{I,3}\\
-\tilde{\Phi}_{I,1}-\tilde{\Phi}_{I,2}-\tilde{\Phi}_{I,3}+3\tilde{\Phi}_{I,4}\\
\vdots\\
\tilde{\Phi}_{I,1}+\tilde{\Phi}_{I,2}+\tilde{\Phi}_{I,3}+\tilde{\Phi}_{I,4}+\cdots-(N-1)\tilde{\Phi}_{I,N}
\end{array}\right)
\end{eqnarray}

\vspace{0.2cm}

\begin{eqnarray}
\Leftrightarrow\ 
\left(\begin{array}{cccccc}
1 & 1 & 1 & 1 & \cdots & 1\\
1 & -1 & 0 & 0 & \cdots & 0\\
1 & 1 & -2 & 0 & \cdots & 0\\
1 & 1 & 1 & -3 & \cdots & 0\\
\vdots & \vdots & \vdots & \vdots & \ddots & \vdots\\
1 & 1 & 1 & 1 & \cdots & -(N-1)
\end{array}\right)\left(\begin{array}{c}
\tilde{\Phi}_{O,1}\\
\tilde{\Phi}_{O,2}\\
\tilde{\Phi}_{O,3}\\
\tilde{\Phi}_{O,4}\\
\vdots\\
\tilde{\Phi}_{O,N}
\end{array}\right) & = & \left(\begin{array}{cccccc}
1 & 1 & 1 & 1 & \cdots & 1\\
-1 & 1 & 0 & 0 & \cdots & 0\\
-1 & -1 & 2 & 0 & \cdots & 0\\
-1 & -1 & -1 & 3 & \cdots & 0\\
\vdots & \vdots & \vdots & \vdots & \ddots & \vdots\\
-1 & -1 & -1 & -1 & \cdots & (N-1)
\end{array}\right)\left(\begin{array}{c}
\tilde{\Phi}_{I,1}\\
\tilde{\Phi}_{I,2}\\
\tilde{\Phi}_{I,3}\\
\tilde{\Phi}_{I,4}\\
\vdots\\
\tilde{\Phi}_{I,N}
\end{array}\right).
\end{eqnarray}

\vspace{0.2cm}

It follows that

\vspace{0.2cm}

\begin{equation}
\left(\begin{array}{c}
\tilde{\Phi}_{O,1}\\
\tilde{\Phi}_{O,2}\\
\tilde{\Phi}_{O,3}\\
\tilde{\Phi}_{O,4}\\
\vdots\\
\tilde{\Phi}_{O,N}
\end{array}\right)=  \underbrace{\left(\begin{array}{cccccc}
\frac{1}{N} & \frac{1}{2} & \frac{1}{6} & \frac{1}{12} & \cdots & \frac{1}{N(N-1)}\\
\frac{1}{N} & -\frac{1}{2} & \frac{1}{6} & \frac{1}{12} & \cdots & \frac{1}{N(N-1)}\\
\frac{1}{N} & 0 & -\frac{1}{3} & \frac{1}{12} & \cdots & \frac{1}{N(N-1)}\\
\frac{1}{N} & 0 & 0 & -\frac{1}{4} & \cdots & \frac{1}{N(N-1)}\\
\vdots & \vdots & \vdots & \vdots & \ddots & \vdots\\
\frac{1}{N} & 0 & 0 & 0 & \cdots & -\frac{1}{N}
\end{array}\right)\left(\begin{array}{cccccc}
1 & 1 & 1 & \cdots & \cdots & 1\\
-1 & 1 & 0 & 0 & \cdots & 0\\
-1 & -1 & 2 & 0 & \cdots & 0\\
-1 & -1 & -1 & 3 & \cdots & 0\\
\vdots & \vdots & \vdots & \vdots & \ddots & \vdots\\
-1 & -1 & -1 & -1 & \cdots & (N-1)
\end{array}\right) }_{\large{=\tilde{\mathbb{M}}}} \left(\begin{array}{c}
\tilde{\Phi}_{I,1}\\
\tilde{\Phi}_{I,2}\\
\tilde{\Phi}_{I,3}\\
\tilde{\Phi}_{I,4}\\
\vdots\\
\tilde{\Phi}_{I,N}
\end{array}\right).
\end{equation}

 \vspace{0.5cm}
Thus the splitting matrix $\tilde{\mathbb{M}} $ for the topological Kondo effect is

\vspace{0.3cm}

\begin{equation}
\tilde{\mathbb{M}} =\left(\begin{array}{cccccc}
2/N -1  & 2/N&  \cdots & 2/N\\
2/N & 2/N -1  &\cdots &2/N\\
\vdots & \vdots  & \ddots & \vdots\\
2/N & 2/N& \cdots& 2/N -1 
\end{array}\right).
\end{equation}

\end{widetext}

For Fermi-liquid leads ($K=1$, i.e.~$\tilde{\mathbb{M}}=\mathbb{M}$), this agrees \cite{beri2,altland1} with the expression $G_{ij} = (e^2/h)(\delta_{ij} - \mathbb{M}_{ij})$ for the $K=1$ conductance tensor.

Hence, according to Eq.~(\ref{delta}), the scaling dimension for electron tunneling into a lead, close to the junction, is 

\begin{eqnarray}
\Delta_i &=& \{ (1-\tilde{\mathbb{M}}_{ii} ) K+(1+\tilde{\mathbb{M}}_{ii} ) /K \}/2  \nonumber \\
& =&  (N-1)K/N + 1/(NK).
\end{eqnarray}

\end{document}